\documentstyle[12pt]{article}
\begin{document}
\newcommand{\scr}{\sin^2 \hat{\theta}_W (m_Z)}
\newcommand{\sef}{\sin^2 \theta_{eff}^{lept}}
\newcommand{\smallms}{{\scriptscriptstyle \rm MS}}
\newcommand{\smallmsbar}{\overline{\smallms}}
\newcommand{\msbar}{\rm{\overline{MS}}}
\newcommand{\be}{\begin{eqnarray}}
\newcommand{\en}{\end{eqnarray}}
\newcommand{\mms}{\hat{m_t}(M_t)}
\newcommand{\mmss}{\hat{m_t}^2(M_t)}
\newcommand{\mpo}{M_t}
\newcommand{\scar}{\hat{s}}
\newcommand{\cc}{\hat{c}}
\newcommand{\as}{\alpha_s}
\newcommand{\amzc}{\hat{\alpha}(m_Z)}
\newcommand{\arun}{\alpha_{\rm run}}
\newcommand{\alc}{\hat{\alpha}}
\newcommand{\nn}{\noindent}
\newcommand{\ew}{electroweak\ }
\newcommand{\ewc}{electroweak corrections\ }
\newcommand{\dres}{(\Delta r)_{\rm res}}
\newcommand{\dr}{\Delta \rho}
\newcommand{\drf}{(\Delta \rho)_f}
\newcommand{\PL}{Phys. Lett.\ }
\newcommand{\NP}{Nucl. Phys.\ }
\newcommand{\PR}{Phys. Rev.\ }
\begin{titlepage}
\begin{flushright}
 NYU--TH--94/08/01 \\
 August 1994 \\
\end{flushright}
\vspace*{2cm}

\centerline{\large{\bf Considerations Concerning the QCD Corrections
 to $\Delta\rho$ }}


\vspace*{1.6cm}

\centerline{\sc Alberto Sirlin.}

\vspace*{.9cm}

\centerline{ Department of Physics, New York University, 4 Washington
Place,}
\centerline{ New York, NY 10003, USA.}

\vspace*{1.2cm}

\begin{center}
\parbox{14.6cm}
{\begin{center} ABSTRACT \end{center}
\vspace*{0.2cm}
Using recent results of Avdeev et al. and an expansion for $\mu_t/\mms$
($M_t$ is the pole mass and $\mu_t\equiv \hat{m_t}(\mu_t)$ ),
it is shown that when $\dr$  is expressed in terms of $\hat{m_t}^2(M_t)$,
the QCD correction is only $(2-3)\times 10^{-3}$ in the NLO approximation.
As a consequence, in terms of $M_t^2$ the  correction to $\dr$ is almost
entirely contained in $\mmss/M_t^2$, a pure QCD effect.
 The latter is studied using various
optimization procedures, and the results compared with the
 expansion proposed
 by Avdeev et al.. Implications for \ew physics are discussed. Threshold
effects are analyzed on the basis of a simple sum rule.
}

\end{center}
\end{titlepage}
\newpage
\nn
The question of the QCD corrections to $\dr$ has been the subject
of several recent studies [1-4].
In particular, Avdeev, Fleischer, Mikhailov, and Tarasov carried out
a complete three-loop calculation of $O(\alpha\as^2)$. Their result,
obtained in the limit $m_b\to 0$, can be expressed to good accuracy
in the form
\be
(\dr)_f= \frac{3G_\mu M_t^2}{8\sqrt{2} \pi^2}\  [1 + \delta_{QCD}],
\en
where
$f$ denotes the fermionic contributions, $\mpo$ is the pole mass, and
\be
\delta_{QCD}= - 2.860 \frac{\as(\mpo)}{\pi} - 10.55 \left( \frac{\as(\mpo
)}{\pi}
\right)^2
\label{avd}
\en
is the QCD correction.
As an illustration, for $\mpo= 200$GeV, Eq.(\ref{avd}) gives $\delta_{QCD}
= -0.0961 - 0.0119= - 0.1080$. Thus, the $O(\as^2)$ term in Eq.(\ref{avd})
is reasonably small and leads to an enhancement of $0.0119/0.0961 = 12.4\%$
of the leading QCD result. Moreover,
 if $\as(\mpo)$ in Eq.(\ref{avd}) is expressed in terms of $\as(\mu)$
and the resulting series is truncated in $O(\as^2(\mu))$, the
$\mu$-dependence
of $\delta_{QCD}(\mu)$ is relatively mild.
For example, for $\mpo=200$GeV and $0.1\le \mu/\mpo\le 1$,
we find a variation $\le 5.8 \times 10^{-3}$, which amounts to $5.4\%$
of the total QCD correction or $49\%$ of the $O(\as^2)$ term.

In spite of these facts, it is worth noting that the expansion
 in Eq.(\ref{avd})
involves large and increasing coefficients, a feature that is frequently
an indication of significant higher order terms.
Furthermore, the arguments of Ref.\cite{2} suggest that there are  at least
two scales in this problem: one, of $O(\mpo)$, associated with
the correction to the \ew amplitude, and another one, much smaller,
 related to  contributions to the pole mass $\mpo$
involving small gluon momenta. It should also be observed
that the $\mu$-sensitivity of the truncated series depends on the
chosen interval, and becomes sharper for $\mu/\mpo <0.1$.
For these reasons, it is a good idea to find alternative expressions for
$\delta_{QCD}$ that separate the two scales, and at the same time involve
terms of $O(\as^2)$ with small coefficients. A simple way of implementing
this idea has been outlined in Ref.\cite{3}.
One expresses first $(\dr)_f$ in terms of $\mms$, the running mass evaluated
at the pole mass, and then relates $\mms$ to $\mpo$ by optimizing the
expansion of $\mpo/\mms$, which is known through $O(\as^2)$ \cite{6}.
As shown below, the arguments of Ref.\cite{3} can
be significantly refined using the new results of Ref.\cite{4}.

Calling $\mu_t$ the solution of
$\hat{m_t}(\mu)= \mu$  and using Eq.(19) of Ref.\cite{4}, $(\dr)_f$ can be
written in the form
\be
\drf =  \frac{3G_\mu \mu_t^2}{8\sqrt{2} \pi^2}\  [1 + \delta_{QCD}^{\msbar}],
\en

\be
\delta_{QCD}^{\msbar}= - 0.19325 \ \frac{\as(\mu_t)}{\pi} +0.07111 \left(
 \frac{\as(\mu_t)}{\pi}\right)^2.
\label{ms}
\en
 We see that the convergence pattern of Eq.(4) is
very nice, with  very small and decreasing coefficients and
 alternating signs.
 For this reason we will assume that the terms of $O(\as^3)$
 and higher are negligible and evaluate Eq.(\ref{ms}) with $\mu_t\to \mpo$
as this introduces only a small change of $O(\as^3)$.
This argument is further strengthened in Appendix A,
where we  discuss the potential contribution of threshold effects
to  perturbative expansions, such as Eq.(\ref{ms}), on the basis of a
simple sum rule.

In order to express $\drf$ in terms of $\mms$, we  use  the following
 next-to-leading order (NLO) expansion, derived in Appendix B:
\be
\frac{\mu_t}{\mms}= 1 + \frac{8}{3} \left(\frac{\as(\mpo)}{\pi}
\right)^2 + [36.33 - 2.45 n_f]\left(\frac{\as(\mpo)}{\pi}\right)^3,
\label{five}
\en
where $n_f=5$ is the number of light flavors.
Writing
\be
\drf= \frac{3G_\mu \mmss}{8\sqrt{2}  \pi^2} \ [1 + \Delta_{QCD}],
\label{sei}
\en
it is possible to combine Eq.(3-5) in  a single expansion
\be
\Delta_{QCD}= -0.19325\  \frac{\as(\mpo)}{\pi} +C \left(
\frac{\as(\mpo)}{\pi}\right)^2.
\label{sette}
\en
In fact, Eq.(\ref{sette}) was proposed in Ref.\cite{3} before the results of
 Ref.\cite{4} were known, and it was argued, on the basis of convergence
assumptions, that $|C|\le6$.
\ From Eqs.(3-5) we see that $C=5.40$, consistent with the arguments
of Ref.\cite{3}. However, with $C=5.40$ the two terms in Eq.(\ref{sette})
nearly cancel and its convergence properties become problematic.
In order to evaluate $\Delta_{QCD}$, it is better to employ the separate
expansions in Eqs.(3,4) and Eq.(\ref{five}). In the latter we retain the
relatively large $O((\hat{\as}/\pi)^3)$ term in order to control the scale
of the leading contribution. Applying the Brodsky-Lepage-Mackenzie (BLM)
optimization \cite{7} to Eq.(\ref{five}), we have
\be
\frac{\mu_t}{\mms}= 1
 + \frac{8}{3} \left(\frac{\as(0.252\mpo)}{\pi}
\right)^2 - 4.10\left(\frac{\as(0.252\mpo)}{\pi}\right)^3 \ \ \ (BLM).
\label{otto}
\en
For $\mpo=200$GeV, Eqs.(\ref{five}) and (\ref{otto}) give $\mu_t/\mms=
1.00392$ and 1.00425, respectively (in this paper $\as(\mu)$ is evaluated with
a 3-loop $\beta$-function for $n_f=5$ light flavors, normalized such
that $\as(m_Z)=0.118$, with $m_Z=91.19$GeV).
We have verified that the evaluations
of  $\mu_t/\mms$ by the priciple of minimal sensitivity
(PMS)\cite{8} and the method of fastest apparent convergence (FAC)\cite{9}
are very close to  the BLM result (the difference is $\le5\times 10^{-6}$).
The correction $\Delta_{QCD}$ in Eq.(\ref{sei}) is obtained from the relation
\be
1+\Delta_{QCD}= \left( \frac{\mu_t}{\mms}\right)^2 \left[
1+ \delta_{QCD}^{\msbar}\right],
\label{nine}
\en
where $\delta_{QCD}^{\msbar}$ and $\mu_t/\mms$ are evaluated via
Eq.(\ref{ms}) (with $\mu_t\to\mpo$) and Eq.(\ref{otto}), respectively.

Table 1 shows that $\delta_{QCD}^{\msbar}$ and $\Delta_{QCD}$
are indeed small corrections for $\mpo\ge 130$GeV. In particular,
$\Delta_{QCD}=(2- 3)\times 10^{-3}$, depending on $\mpo$.
This conforms with the idea
that $\mms$ is a good expansion parameter in the sense that the associated
QCD corrections are small \cite{3}.
Furthermore,  in the $\drf$ case
this occurs to a rather remarkable degree.
However,
it is not a good approximation to retain only the $O(\as)$
correction in Eq.(6), as it is nearly cancelled by higher order contributions.
On the other hand, the expansion in Eq.(\ref{ms}) is very well behaved.
In this sense, $\mu_t$ is a better expansion parameter than $\mms$, which
may reflect the fact that it makes no reference to the pole mass
$\mpo.$\footnote{The usefulness of $\mu_t$ as
 an expansion parameter has also been emphasized in Ref.\cite{buras}.}

Following the strategy outlined in Ref.\cite{3},
$\delta_{QCD}$ in Eq.(1) can be obtained from the relation
\be
1 +\delta_{QCD}=
\left(\frac{\mms}{\mpo}\right)^2 [1+ \Delta_{QCD}].
\label{ten}
\en
For $\mpo/\mms$ we have the well-known expansion \cite{6}
\be
\frac{\mpo}{\mms}=
1+\frac{4}{3} \frac{\as(\mpo)}{\pi} +(16.11 - 1.04 n_f) \left(
\frac{\as(\mpo)}{\pi}\right)^2,
\label{undici}
\en
where the masses of the light quarks have been neglected.
As $\mpo/\mms$ plays a central role in our final result, we record
the expressions obtained by applying the three optimization methods
to the r.h.s. of Eq.(\ref{undici})
\be
\frac{\mpo}{\mms}=
1 +\frac{4}{3} \frac{\as(\mu^*)}{\pi} - 1.05 \left(
\frac{\as(\mu^*)}{\pi}\right)^2 \ \ \ (BLM),
\label{12}
\en
\be
\frac{\mpo}{\mms}=
1 +\frac{4}{3} \frac{\as(\mu^{**})}{\pi} - 0.84 \left(
\frac{\as(\mu^{**})}{\pi}\right)^2 \ \ \ (PMS),
\label{13}
\en
\be
\frac{\mpo}{\mms}=
1 +\frac{4}{3} \frac{\as(\mu^{***})}{\pi}\ \ \ \ \  \ \ \ \ \ \ (FAC),
\label{14}
\en
where $\mu^*= 0.0963 \mpo$\cite{3},
$\mu^{**}= 0.1004\mpo$, $\mu^{***}=0.1183\mpo$.
We note that  Eqs.(12-14) involve coefficients of $O(1)$ and similar scales.
For $\mpo=200$GeV, Eqs.(12-14) give $\mpo/\mms=1.06305$, 1.06304,
and 1.06299, respectively. For $\mpo=174$GeV, the corresponding values are
1.06479, 1.06478, 1.06472.
Finally, for $\mpo= 130$GeV, we have 1.06877, 1.06876, 1.06869.
Thus, the three approaches give remarkably consistent results.
In contrast, the expansion in Eq.(\ref{undici}), which involves
a large second order coefficient, gives, for $\mpo=(200,174,130)$GeV,
1.0571, 1.0584, 1.0614, respectively, which are $(0.6- 0.7)\%$ smaller.
One must conclude that the coefficients
of the unknown terms of $O((\as/\pi)^3)$ and higher in Eq.(\ref{undici})
and or Eqs.(12-14) are large. For instance, if the truncated BLM
expansion in Eq.(\ref{12}) were exact, the coefficient of the
$(\as(\mpo)/\pi)^3$  and $(\as(\mpo)/\pi)^4 $  terms in Eq.(\ref{undici})
would be $\approx$\ 104 and \ 1,041, respectively.\footnote{
For a recent application of the PMS and FAC approaches to estimate higher
order coefficients in other cases, see Ref.\cite{kat}.}
 In the following we employ
the optimized expression for $\mpo/\mms$, which, for definiteness, we
identify with Eq.(\ref{12}).

Table 2 displays the values of $\delta_{QCD}$ obtained from Eq.(\ref{ten}),
using Eq.(\ref{12}) and our previous determination of $\Delta_{QCD}$
(Eq.(\ref{nine})), and compares them with those derived from Eq.(\ref{avd}).
In order to show the effect of the higher order contributions (H.O.C.),
 we also exhibit the
fractional enhancement of the total QCD correction over the conventional
$O(\as)$ result (first term in Eq.(\ref{avd})).

As pointed out by Kniehl \cite{12}, the various evaluations of QCD
corrections can be conveniently interpreted as
effective re-scalings of the $O(\as)$ result.
In fact, we find that
the two evaluations of $\delta_{QCD}$  in Table 2 can be very well
 represented by the simple formulas
\be
\delta_{QCD}= - 2.86 \ \frac{\as(0.444\mpo)}{\pi} \ \ \ [Eq.(\ref{avd})],
\label{15}
\en
\be
\delta_{QCD}= - 2.86 \ \frac{\as(0.323\mpo)}{\pi} \ \ \ [Eq.(\ref{ten})].
\label{16}
\en
 The effective scales have been  chosen
so that Eqs.(\ref{15},\ref{16}) reproduce the values in Table 2
with an error of at most 1 or 2 times $10^{-4}$.

\ From Table 2 we see that, for 130GeV$\le\mpo\le$220GeV,
the evaluation of $|\delta_{QCD}|$ based on Eq.(\ref{ten})
is $(5.1- 6.5)\times 10^{-3}$ larger than the results from Eq.(\ref{avd}).
 This difference amounts to $\approx 5\%$ of the total QCD correction.
On the other hand, the last two columns in Table 2 show that the H.O.C.
 are  $\approx 45\%$ larger in the evaluation based
on Eq.(\ref{ten}). As the two calculations coincide through terms of
$O((\as(\mpo)/\pi)^2)$, one must conclude that the coefficients of the
$O((\as/\pi)^3)$ and higher terms in Eq.(\ref{avd}) and or Eq.(\ref{ten})
are large. For instance, if Eq.(\ref{ten}) (with $\Delta_{QCD}$ and
 $\mpo/\mms$
evaluated via Eqs.(\ref{nine}) and ( \ref{12}))
were exact, the coefficient of the neglected $(\as(\mpo)/\pi)^3$ term in
Eq.(\ref{avd}) would be $\approx -93.4$.
Of course, it could happen that the exact answer
lies between the two evaluations, in which case the $O(\as^3)$ terms
would be smaller.
 However, because of its pattern of large and increasing
coefficients, we think that
this problem is more likely to occur in Eq.(\ref{avd}).
It is also useful to note that if one applies the optimization
procedures to Eq.(\ref{avd}), the results become closer to those from
Eq.(\ref{ten}). Specifically, we have

\be
\delta_{QCD}= - 2.86 \frac{\as(0.154\mpo)}{\pi} +
9.99 \left(\frac{\as(0.154\mpo)}{\pi} \right)^2\ \ \ (Eq.(\ref{avd});\ BML),
\label{17}
\en
\be
\delta_{QCD}= - 2.86 \frac{\as(0.324\mpo)}{\pi} +
1.80 \left(\frac{\as(0.324\mpo)}{\pi} \right)^2\ \ \ (Eq.(\ref{avd});\ PMS),
\label{18}
\en
\be
\delta_{QCD}= - 2.86 \ \frac{\as(0.382\mpo)}{\pi}
\ \ \ \ \ \ \ \ (Eq.(\ref{avd});\ FAC).
\label{19}
\en
For $\mpo=200$GeV, Eqs.(\ref{17}-\ref{19}) give 0.1085,
0.11044, 0.11037, respectively. We see that the difference between Eqs.(
\ref{18},\ref{19}) and Eq.(\ref{ten}) is $\approx 45\%$ smaller than that
between Eq.(\ref{avd}) and Eq.(\ref{ten}).
It has also been pointed out that in most NLO QCD calculations, it is a good
approximation to retain only the leading term, evaluated at the BLM scale
[2,7].
\ From Eq.(\ref{17}) we see that this is not the case for $\drf$, as
the coefficient of the residual $O((\as/\pi)^2)$ term, 9.99, is quite large.

There is a simple but important point
that should be stressed: once it is recognized that $\delta_{QCD}$ is almost
entirely contained in the first factor of Eq.(\ref{ten}) (and this follows
from Eq.(4) and (8),  which are  quite robust), it becomes
clear that its magnitude and precision are largely controlled
by the value of $(\mpo/\mms)^2$ and the accuracy within which
it can be calculated.
 We emphasize that this is a pure QCD, rather than \ew effect.
 We have also pointed out that there is a
significant difference between the evaluations of this correction
via Eq.(\ref{undici}) or by means of Eqs.(\ref{12}-\ref{14}).
In this paper we have taken the point of view that $(\mpo /\mms)^2$
can be best evaluated at present by means of the optimized expansions of
 Eqs.(\ref{12}-\ref{14}). In particular,
we have noted that the coefficients in these equations are $O(1)$
and that the
values derived from the three optimization procedures remarkably
consistent. In fact, it would be very interesting to check this approach
by an explicit evaluation of the unknown third order term in
Eq.(\ref{undici}).

As shown in Table 2, the formulation of this paper, based in
Eqs.(\ref{nine},\ref{ten},\ref{12}) and summarized in the effective formula
 of Eq.(\ref{16}), leads, for $\mpo=200$GeV, to an enhancement of
18.0$\%$ relative to the conventional $O(\as)$ calculation.
This modification corresponds to an additional contribution
$-2.2\times10^{-4}$ to $\drf$ and $+0.76\times 10^{-3}$ to $\Delta r$,
and shifts the predicted values of $\mpo$ and $m_W$ by
$+1.9$GeV and $-13$MeV, respectively.
It can be compared with an enhancement  of $12.4\%$ from Eq.(\ref{avd}),
or 14.9$\%$ from the optimized expressions of Eq.(\ref{18}) or Eq.(\ref{19}).
Although the H.O.C. obtained from Eq.(\ref{avd}) and
 Eq.(\ref{ten}) are significantly different,  the effects
of this difference on \ew physics, namely  variations of 0.6GeV in
$\mpo$ and 4MeV in $m_W$, are rather small.
As mentioned before, it is possible that the exact answer lies
between the two calculations.
 Although we prefer the evaluation based on Eq.(10), the difference with
Eq.(2) is interesting because it may be used as an estimate of the theoretical
error due to uncalculated higher order terms. We recall that there is
at present  an
additional uncertainty of $\approx 5\%$ due to the  $\pm0.006$ error
in $\as(m_Z)$\cite{3}. Following the discussion of Ref.\cite{3}, it is also
easy to see how the approach in this paper can be applied to other \ew
amplitudes proportional to $\mpo^2$, such as those present in
$Z^0\to b\bar b$.

Finally, it is instructive to compare the above results  with some of the
estimates made before the complete $O(\as^2)$ contribution became known.
Although conceptually very different, the dispersive approach as implemented
in the last paper of Ref.\cite{1}, gives, for 130GeV$\le\mpo\le$220GeV,
results very close numerically to those of Ref.\cite{2}.
The latter approximates the relevant Feynman diagrams by the $O(\as)$
contribution evaluated at the BLM scale (first term of Eq.(17)).
For $\mpo=200$GeV, both estimates led approximately
to a 34$\%$ enhancement of the $O(\as)$
contribution, a result of the same sign but clearly larger than the
conclusions from either Eq.(2) or Eq.(10). On the other hand, out previous
estimate $[(26\pm 6)\%]$ \cite{3}, is roughly consistent with Eq.(10),
but larger than the values from Eq.(2).

\vskip 1.3cm
\noindent{\bf{Aknowledgements}}
\vskip .4cm
The author is indebted to S. Brodsky, J. Fleischer, P. Gambino, A. Kataev,
B. Kniehl,  and W. Marciano for  useful
discussions.
This work was supported in part by the National Science Foundation under
 Grant No. PHY--9313781.
\vskip 1.5cm

\subsection*{APPENDIX A}
\renewcommand{\theequation}{A.\arabic{equation}}
\setcounter{equation}{0}

We briefly discuss threshold effects on the basis of a simple sum rule that
 follows as a corollary from the arguments of Ref.\cite{13}.
In that work  the operator product expansion
\cite{14} is applied to investigate the validity  of
 the dispersion relations [D.R.] for vacuum polarization functions
previously proposed by Chang et al. and Kniehl and Sirlin \cite{1},
 to all orders in perturbation
 theory. We call $\Pi^{V,A}_{\mu\nu}(q,m_1,m_2)$ the vacuum polarization
tensors associated with the vector and axial vector currents attached
to pairs of quarks, such as $t\bar t, \ t\bar b, b \bar b$, endowed with
masses $m_1 $ and $m_2$. Decomposing
\be
\Pi^{V,A}_{\mu\nu}(q,m_1,m_2 )=
\Pi^{V,A}(q^2,m_1,m_2) g_{\mu\nu} + \lambda^{V,A}(q^2,m_1,m_2)q_\mu q_\nu,
\en
we have the sum rule
\be
\int^\infty ds'& Im&\left[\sum_{i=1}^2 \frac{1}{2}\left(
\lambda^V(s',m_i, m_i)+\lambda^A(s',m_i,m_i)\right)\right.\nonumber\\
&&\left.-\lambda^V(s',m_1,m_2)-\lambda^A(s',m_1,m_2)\right]=0.
\en
In order to derive Eq.(A.2), one invokes analiticity and applies
Cauchy's theorem to the appropriate combination of $\lambda$'s
over a contour with straight lines just above and below the positive real
axis, closed by a large circle.
The arguments of Ref.\cite{13} and the results of
 Ref.\cite{14} then show that the contribution of the large circle
vanishes in the limit of infinite radius.
Identifying $m_1=m_t$, $m_2=m_b$, there are large and equal non-relativistic
 contributions
from the  threshold region to $Im\lambda^V
(s',m_t,m_t)$ and $Im\lambda^A(s',m_t,m_t)$ (threshold contributions to
 the other $\lambda$'s are suppressed by reduced mass effects).
However, the appropriate D.R.\cite{1} involve
the \ integrals \ $\int^\infty ds' Im\lambda^A (s',m_t,m_t)$
or \  $\int^\infty ds' Im\lambda^V (s',m_t,m_t)$
over all values of $s'$.
It is precisely from the contributions of such integrals in the D.R.
that  large threshold effects may potentially arise.
On the other hand, Eq.(A.2) tells us that
$\int^\infty ds' Im[\lambda^V (s',m_t,m_t)+\lambda^A(s',m_t,m_t)]$
can be expressed as a linear combination of integrals of the spectral
 functions
$Im \lambda^{(j)}(s',m_t,m_b)$
and $Im \lambda^{(j)}(s',m_b,m_b)$ $(j=V,A)$  which do not have significant
threshold contributions and involve different channels.
The most sensible interpretation of this result is that
the large threshold contributions to
 $Im\lambda^V (s',m_t,m_t)$ and  $\lambda^A(s',m_t,m_t)$
 cancel against other contributions when the integrals over all
 $s'$  values are considered. As it is known that such integrals can be
expanded perturbatively in powers of $\as$, this cancellation must occur
order by order in pertubation theory.
 In particular, this applies to the
expansion in Eq.(4). Similar conclusions were reached in Ref.\cite{2}
on the basis of different arguments.
In summary, although the sum rule in Eq.(A.2) does not provide a rigorous
 proof of the cancellation of large threshold effects in \ew
amplitudes, it gives strong support to such conclusion when all
contributions of a given order are included.
On the other hand, it does not reveal the detailed mechanism
of the cancellation,
 an interesting problem which hopefully can be clarified
in the future.

\subsection*{APPENDIX B}
\renewcommand{\theequation}{B.\arabic{equation}}
\setcounter{equation}{0}

In order to derive Eq.(5), we start with the well-known expression
\cite{14}
\be
\hat{m_t}(\mu)= m_t^* \left(-\beta_1\frac{\as(\mu)}{\pi}\right)^d
\left[1+a_1 \frac{\as(\mu)}{\pi} + a_2 \left(\frac{\as(\mu)}{\pi}
\right)^2 + ...\right],
\en
where $m_t^*$ is the RG invariant mass,
$\beta_1=\frac{n_f}{3}-\frac{11}{2}$, $d=-2/\beta_1$, $a_1=(\beta_2/\beta_1)
[2/\beta_1 - \gamma_2/\beta_2]$,
$\gamma_2= \frac{101}{12} - \frac{5n_f}{18}$
(the other quantities need not be specified for our purposes).
Setting $\mu=\mu_t\equiv\hat{m_t}(\mu_t)$
in Eq.(B.1) and dividing by $\mms$ we have
\be
\frac{\mu_t}{\mms}=
\left(\frac{\as(\mu_t)}{\as(\mpo)}\right)^d
\frac{\left[1+a_1\frac{\as(\mu_t)}{\pi} + a_2 \left(\frac{\as(\mu_t)}{\pi}
\right)^2 +...\right]}
{\left[1+a_1\frac{\as(\mpo)}{\pi} + a_2 \left(\frac{\as(\mpo)}{\pi}
\right)^2 +...\right]}
\en
In the numerators we expand $\as(\mu)$ in powers of $\as(\mpo)$. The first
factor gives
\be
\left(\frac{\as(\mu_t)}{\as(\mpo)}\right)^d =
1+ d \ \frac{\as(\mpo)}{\pi} \ln\left(\frac{\mu_t}{\mpo}\right)
\left(\beta_1 + \beta_2 \frac{\as(\mpo)}{\pi}\right) + O(\as^4)
\en
where we have used the fact that $\ln (\mu_t/\mpo) = O(\as)$.
Similarly, we see that the second factor in Eq.(B.2)
becomes $1+a_1 \beta_1 (\as(\mpo)/\pi)^2 \ln(\mu_t/\mpo) + O(\as^4)$.
Combining the two factors, noting that $\beta_1 d= -2$,
$\beta_2 d+ a_1\beta_1=-\gamma_2$, and splitting the logarithm we have
\be
\frac{\mu_t}{\mms}= 1-
\frac{\as(\mpo)}{\pi}\left[
\ln\left(\frac{\mu_t}{\mms}\right)+
\ln\left(\frac{\mms}{\mpo}\right)\right]\left[
2+\gamma_2 \frac{\as(\mpo)}{\pi}\right]+O(\as^4)\nonumber\\
\en
Inserting $\mu_t/\mms = 1+\sum_{n=1}^{3}x_n (\as(\mpo)/\pi)^n$
in both sides of Eq.(B.4), using Eq.(11) in the second logarithm
and matching coefficients,
we obtain $x_1=0$, $x_2=8/3$, $x_3=2(16.11-1.04 n_f - 8/9) +
 4 \gamma_2/3 - 2
x_2= 36.33 - 2.45 n_f$,
which leads to Eq.(5).
This derivation assumes that $\hat{m_t}(\mu)$ evolves according to the $n_f$
light quarks. The difference in $\gamma_2$ and $x_3$ when one employs
6 active quarks, as in Ref.\cite{4}, is very small.

\newpage
\begin{center}
$$
\begin{array}{|r|r|r|}\hline
  \mpo ({\rm GeV}) & \ 10^3 \delta_{QCD}^{\msbar}\
 & 10^3 \Delta_{QCD} \\ \hline
     130  & -6.80 & 2.98\\ \hline

   150 &-6.67& 2.65 \\ \hline

   174& -6.53& 2.33\\ \hline

   200& -6.41& 2.06 \\ \hline

 220& -6.33& 1.88 \\ \hline
   \end{array}
$$
\end{center}
\vskip 0.5cm
{\bf Table 1.} The corrections
$\delta_{QCD}^{\msbar}$ and $\Delta_{QCD}$. The first one is given by Eq.(4)
with $\mu_t\to\mpo$, while the second is obtained from Eq.(9),
with $\mu_t/\mms$ evaluated according to Eq.(8) ($\as(m_Z)=0.118$
 is employed).

\vskip 1.7cm
\begin{center}
$$
\begin{array}{|r|r|r|r|r|}\hline
  \mpo \ \  & \  \delta_{QCD}\  &  \ \delta_{QCD}\  & H.O.C.
& H.O.C. \\

  ({\rm GeV}) & ({\rm Eq.(2)}) & ({\rm Eq.(10)})& ({\rm Eq.(2)})\% &
({\rm Eq.(10)})\% \\ \hline

130  & -0.1154 & -0.1219 & 13.2 & 19.5\\ \hline

   150 &-0.1128& -0.1189 & 12.9 & 19.0\\ \hline

   174& -0.1102& -0.1159 & 12.6 & 18.4\\ \hline

   200& -0.1080& -0.1133 & 12.4 & 18.0\\ \hline

 220& -0.1064& - 0.1115 & 12.2 & 17.6\\ \hline
   \end{array}
$$
\end{center}
\vskip 0.5cm
{\bf Table 2.} Comparison of two determinations of $\delta_{QCD}$.
The second column is based on Eq.(2) [4]. The third column is based on
Eq.(10) with $\mpo/\mms$ obtained from Eq.(12) and $\Delta_{QCD}$ evaluated
according to Table 1. The fourth and fifth columns give the fractional
enhancement over the conventional $O(\as)$ result (first term
of Eq.(2)) due to the inclusion of higher order contributions (H.O.C.).

\end{document}